\documentclass[11pt,a4,fleqn]{article}
\usepackage{graphicx}
\usepackage{amsmath,amssymb,latexsym,graphics,epsfig}
\usepackage{color}
\usepackage{amsthm}

\usepackage{algorithm}
\usepackage{algpseudocode}
\usepackage{setspace}
\usepackage{caption}

\DeclareMathOperator{\cost}{cost}

\numberwithin{equation}{section}

\begin{document}

\title{An Efficient Parallel Data Clustering Algorithm Using Isoperimetric Number of Trees}

\author{\small  Ramin Javadi$^{\textrm{a}}$, Saleh Ashkboos$^{\textrm{b}}$\\
\small  $^{\textrm{a}}$Department of Mathematical Sciences,
Isfahan University
of Technology,\\ \small Isfahan, 84156-83111, Iran\\
\small  $^{\textrm{b}}$Department of Electrical and Computer Engineering,
Isfahan University
of Technology,\\ \small Isfahan, 84156-83111, Iran\\
\small \texttt{E-mails: rjavadi@cc.iut.ac.ir, s.ashkboos@ec.iut.ac.ir}\\
}
\date {}

\maketitle 

\begin{abstract}
\noindent

We propose a parallel graph-based data clustering algorithm using CUDA
GPU, based on exact clustering of the minimum spanning tree in terms of a minimum isoperimetric criteria. We also provide a comparative performance analysis of our algorithm with
other related ones  which demonstrates the general superiority of this parallel algorithm over other competing algorithms in terms of accuracy and speed.


\end{abstract}

\section{Introduction}\label{sec:introduction}

%
%
%
%

Data clustering as the unsupervised  grouping of similar data which have high intra-similarity and low inter-similarity is one of the most important tasks in many fields  such as data mining, signal and image processing, computational biology, computer vision and etc.

In this way, graph-based methods as deriving a weighted graph from data and applying the graph partitioning algorithms is one of the most efficient approaches which has been studied widely in the literature. For a review on various graph-based clustering approaches, see \cite{jain1999data} and \cite{xu2005survey}.

Shi and Malik \cite{shi2000normalized} introduced the normalized cut problem (Ncut for short) as an appropriate formulation of the clustering criteria. The objective of the normalized cut problem is to partition the vertex set of a graph into $ k $ clusters such that the average of the normalized sparsity of the clusters is minimized. They showed that the decision problem corresponding to the computation of the minimum Ncut of a graph is NP-complete in general and they proposed an approximation procedure (based on spectral techniques) by relaxing the characteristic functions of the clusters into the real domain and using the information provided by the eigenspace of the corresponding Laplacian matrix.   Recently, Trevisan et al. \cite{lee2014multiway}, by proving higher-order Cheeger's inequalities, have provided a theoretical justification for the empirical efficiency of the spectral clustering methods. 

Daneshgar et al. \cite{daneshgar2012complexity,daneshgar2013clustering} proved that the normalized cut problem is NP-complete on trees (even unweighted trees)  and they proposed to relax the search space from all $ k- $partitions to all $ k- $subpartitions (a $ k- $subpartition is a set of $ k $ disjoint subsets of the vertex set which is not necessarily a partition). They achieved a polynomial time algorithm (actually a near linear algorithm) for this relaxed problem (called the isoperimetric problem) on weighted trees and revealed the efficiency of the proposed method applying in the clustering of the real-world data. They also provided some evidence that how the information of residual nodes remained outside the optimal subpartition can be used to extract outlier data points, in the course of performing the clustering process.

In the first step of their algorithm, a weighted graph is created from the data based on some measure of data features. Then, a minimum spanning tree of the graph is computed and the algorithm starts to find $k'$th isoperimetric number of the tree.  The result is an optimal subpartition of the nodes which yields a clustering of the dataset.

Modifying the standard methods in order to let the computations be performed in parallel can remarkably improve the time performance of the algorithms. Accordingly, this issue, along with the hardware and software development of  parallel computation, have been at the center of attention in recent decades.  
In this paper, we develop an efficient parallel data clustering algorithm based on isoperimetric number of trees. Our approach can be considered as a parallel version of Daneshgar et al. algorithm \cite{daneshgar2013clustering}, in which the computational performance is significantly improved. 

In recent decades, many parallel computing architectures have been developed  falling into two basic categories:	 CPU based and GPU (Graphics Processor Unit) based architectures. Unlike the CPUs, GPUs are specially developed for running and executing parallel codes, and this feature makes them more powerful and cheaper for performing the computational goals.
Also, GPUs have very large memory bandwidth and less data transfer overhead in compare with other parallel architectures (Their memory bandwidth are up to 300 GB/s in recent years). 
In addition, GPUs have become easy programmable using a high level language, called CUDA, which have been developed for this purpose. Today, GPUs are not only flexible, but also easy programmable devices and this makes them one of the most popular devices for applied and theoretical  research purposes (see \cite{owens2007survey}).

In order to achieve a better performance, we implemented our algorithm on NVidia GPUs by using CUDA platform. 
The experimental results on real-world datasets reveal a final speed up between 1.36x and 21x. We also show that the efficiency is significantly improved in higher dimensional data such as generated data from social networks.
According to the comparison results, we believe that our algorithm is one of the most efficient data clustering algorithms in the case of speed and accuracy.

\subsection{Related work}

Various algorithms and strategies have been proposed to improve the computational performance of data clustering and its applications to image segmentation, signal processing and etc. In this manner, some attempts have been devoted to develop parallel adaptations of the well-known methods. 
 
K-means is one of the most popular algorithm in data clustering literature. Firstly,  Stoffel and  Belkoniene \cite{stoffel1999parallel} developed some parallel techniques for this algorithm. Afterwards, many attempts was devoted to design and implement the parallel version for k-means in the large data sets (e.g. see \cite{dhillon2000data} and \cite{joshi2003parallel}).

Normalized cut \cite{shi2000normalized} is another popular graph-based data clustering algorithm for which XianLou and ShuangYuan \cite{xianlou2013image} developed a parallel version and gained about 2.34 times speed up in the best case.
In the context of hierarchical clustering, firstly, Olson \cite{olson1995parallel}, introduced a parallel algorithm. Also, Dahlhaus \cite{dahlhaus2000parallel} presented another parallel algorithm and used his algorithm to solve the undirected split decomposition problem in graphs. Amal Elsayed Aboutabl et al. \cite{aboutabl2011novel}, designed and implemented a parallel document clustering based on hierarchical approach and for the best case, they gained a 5 times speed up over $65 \%$ efficiency. 

On the other hand, some other parallel algorithms are developed on the various architectures and models. Weizhong et al.  \cite{zhao2009parallel}, implemented a parallel version of k-means clustering on map-reduce model.  Middelmann and Sanders \cite{wassenberg2009efficient} proposed a heuristic graph-based image segmentation algorithm and implemented a parallel algorithm for shared-memory machines.

Most of these endeavours are suffered from accuracy and time performance (achievable speed-up in compare with the sequential version).

\section{Data clustering: sequential algorithm}

\subsection{Graph-based clustering}
Graph-based data clustering is one of the most important approaches in data clustering. In this approach, the data set is represented as a weighed graph such that each node corresponds to a data point and the weight of each edge between two nodes reflects their proximity. Finally, the clustering problem is reduced to a graph partitioning problem according to a prescribed criteria.\\

In our model, given a data set $ X=\lbrace x_{1},\ldots,x_{n}\rbrace \subseteq \mathbb{R}^d $, the corresponding graph is a weighted graph $ G $ on the vertex set $ X $ and the edge set $ E $ (the set of all unordered pairs of $ X $) with the following weight functions:

\begin{itemize}

  \item An edge-weight function $\varphi$ : $E$ $\rightarrow$ $ \mathbb{R}^{+}$, called the \textit{flow}, that represents the similarities between pairs of elements which is defined as,
  
\begin{equation}
	\forall 1 \leq i \neq j \leq n, \  \varphi(x_{i},x_{j}):= \exp(-\|x_{i}-x_{j}\|_{2}/\sigma ),  \label{phi}
		\end{equation}

where $\sigma$ is a scaling parameter.

\item A vertex-weight function $\omega$ : $X$ $\rightarrow $    $ \mathbb{R}^{+}$, stands for the weight of each data element defined as,
\begin{equation}\label{omega}
\forall 1\leq i \leq n, \qquad  \omega(x_{i}) :=\sum\limits_{j=1}^{n} \varphi(x_{i},x_{j}).
\end{equation}

 \item Another function $p$ : $X$ $\rightarrow$ $ \mathbb{R}^+\cup \{0\}$  called the \textit{potential function} which shows the isolation of each data element defined as,
\begin{equation} \label{potential}
\forall 1\leq i \leq n, \qquad  p(x_{i}) := \alpha\sum\limits_{j=1}^{n} \|x_{i}-x_{j}\|_{2},
\end{equation}
where $\alpha$ is a scaling factor and is assumed to be equal to zero, unless an extra outlier detection tasks is needed (see \cite{daneshgar2013clustering}). 
\end{itemize}

For a real function $f$ on the set $X$ and a subset $A\subset X$, we define $f(A):= \sum_{x\in A} f(x)$.
Also, we denote the \textit{boundary} of $A$ by $\delta(A)$ and define as,

\[	\delta(A) :=\lbrace xy \in E \mid x\in A,y\in X \backslash A\rbrace.\]

The collection of all \textit{$k$-partitions} of $X$ is denoted by $P_{k}(X)$ and defined as,

\begin{align*}
P_{k}(X):=\lbrace &\mathcal{A}=\lbrace A_{1},\ldots,A_{k}  \rbrace \mid \forall i,A_{i}\subset X, \\
&\forall i\neq j,A_{i}\cap A_{j} =\emptyset \text{ and } \bigcup_{i=1}^{k} A_{i} = X  \rbrace.
\end{align*}

Also, the family of all \textit{$k$-subpartitions} of $X$ is denoted by $D_{k}(X)$ and defined as,
\begin{align*}
D_{k}(X) :=\lbrace &\mathcal{A}=\lbrace A_{1},\ldots,A_{k}  \rbrace \mid \forall i,A_{i}\subset X, \\
&\text{ and } \forall i\neq j,A_{i}\cap A_{j} =\emptyset  \rbrace.
\end{align*}

The \textit{$k$-normalized cut problem} is to find a $k$-partition $\mathcal{A}\in P_{k}(X)$ minimizing the following cost function,
	\begin{equation}
				\cost(\mathcal{A}):=\max_{1\leq i\leq k}  \frac{\varphi (\delta A_{i}) + p(A_{i})}{\omega(A_{i})}.\label{cost}
	\end{equation}

Let us define,
		\begin{equation}
				MNC_{k}(G):= \min_{\mathcal{A}\in P_k(X)} \cost(\mathcal{A}). \label{MNC}
		\end{equation}

The \textit{isoperimetric problem} seeks for a minimizer of the cost function in \eqref{cost} over the space of all $k$-subpartitions $D_{k}(X)$. We denote the minimum by $MISO_{k}(G)$ and define as,

		\begin{equation}
		MISO_k(G):= \min_{\mathcal{A}\in D_k(X)} \cost(\mathcal{A}). \label{MISO}
		\end{equation}
	
The quotient ($\varphi(\delta A) + p(A))/\omega(A)$ is called the \textit{normalized sparsity} of the set $A$.

\subsection{Sequential algorithm}
In  \cite{shi2000normalized}, it is shown that the normalized cut problem is NP-hard for general graphs (even when $k = 2$). Also, in \cite{daneshgar2012complexity} it is shown that this problem is NP-hard on weighed-trees when $k$ is given in the input. In contrast, Daneshgar et al. in \cite{daneshgar2013clustering} shows that the isoperimetric problem is efficiently solvable on weighted tree using a near linear time algorithm. The outline of their method can be described as follows.
For a given number $N$, they use Algorithm~\ref{Main Sequential decision Algorithm} to check if there exists a subpartition with the cost at most $N$. Then, in order to find the minimizing subpartition, they apply Algorithm~\ref{Main Sequential Algorithm} which executes a binary search and iteratively calls Algorithm~\ref{Main Sequential decision Algorithm} as a subroutine. In fact, Algorithm~\ref{Main Sequential decision Algorithm} is essentially a backtracking which traverses through the vertices and check if a vertex can be either cut from or join to its parent, according to some sort of conditions.

The detail of Algorithms 1 and 2 are elaborated in the following. To prove the correctness of Algorithms~\ref{Main Sequential decision Algorithm} and \ref{Main Sequential Algorithm}, see \cite{daneshgar2013clustering}.

\begin{algorithm}[ht]
\caption{\\
\textbf{Input:} A weighted tree ($T,\omega,\varphi,p$) rooted at the vertex $v$, an integer $k$
and a rational number $N$. (It is assumed that the vertices are ordered in BFS order as $x_{1},\ldots ,x_{n} = v$.)\\
\textbf{Output:} A $k$-subpartition $\mathcal{A}\in D_{k}(X)$  (if there exists) such that $\cost(\mathcal{A})\leq N$ as in \eqref{cost}.}\label{Main Sequential decision Algorithm}
\begin{algorithmic}[]
\State Initialize the set function $\eta$ : $X$ $\rightarrow$ $P(X)$ by $\eta(x_{i})$ = \{$x_{i}$\} for each $1 \leq i \leq n$.
\State Define $i$ = 1 , $j$ = 0.
\While{$j<k$ and $i\leq n$}
\State Let $u$ be the parent of $x_i$ and $e=x_iu\in E$ be the parent edge.
		\If {$\varphi(e)+p(x_{i}) \leq N\omega(x_{i})$ }
			\State $j\leftarrow j+1$, $A_{j}\leftarrow \eta(x_{i})$, $\omega(A_{j})\leftarrow
			 \omega(x_{i})$
			 \State $\varphi(\delta A_{j}) \leftarrow \varphi(e)+p(x_{i})$, $p(u)\leftarrow 	   								p(u)+\varphi(e)$.
		\ElsIf{$p(x_{i}) - \varphi(e) < N\omega(x_{i})$}
			\State $\eta (u)\leftarrow \eta(u)\cup\eta(x_{i})$, $\omega(u)\leftarrow \omega(u)                       				 +\omega(x_{i})$, $p(u)\leftarrow p(u)+p(x_{i})$.
		\Else
			\State $p(u)\leftarrow p(u)+\varphi(x_{i})$.
		\EndIf
	\State $i\leftarrow i+1$.
\EndWhile
\If {$j$ = $k$}
\State \Return YES and $\mathcal{A}=\lbrace A_{1},\ldots,A_{k} \rbrace$.
\Else
\State \Return NO.
\EndIf
\end{algorithmic}
\end{algorithm}

Now, suppose that $\varphi^{*}$, $\varphi_{*}$, $p^{*}$, $p_{*}$, $\omega^{*}$ and $\omega_{*}$ are defined as follows.
\begin{align*}
\varphi^{*}&:= \sum\limits_{e\in E}\varphi(e), \hspace*{50pt}  &\varphi_{*}:= \min\limits_{e\in E}\varphi(e),\\
\omega^{*}&:= \sum\limits_{x\in X}\omega(x), \hspace*{50pt} &\omega_{*}:= \min\limits_{x\in X}\omega(x), \\
p^{*}&:= \sum\limits_{x\in X}p(x), \hspace*{50pt} & p_{*}:= \min\limits_{x\in X}p(x).
\end{align*}
Algorithm~\ref{Main Sequential Algorithm} finds the minimizing subpartition.
\begin{algorithm}[ht]
\caption{\\
\textbf{Input:} A rooted weighted tree ($T,\omega,\varphi,p$), an integer $k$. \\
\textbf{Output:} A minimizing subpartition achieving $MISO_{k}(T)$.}\label{Main Sequential Algorithm}
\begin{algorithmic}[]
	\State Let $\alpha_{0} \longleftarrow \frac{\varphi_{*} + p_{*}}{\omega^{*}}$ and $\beta_{0}						 \longleftarrow \frac{\varphi^{*} + p^{*}}{\omega_{*}}$.
	\State Let $t \longleftarrow \log(2{\omega^{*}}^{2}(\beta_{0} - \alpha_{0}))- \log(\varphi_{*} + p_{*} )$.
	\State Initialize $\alpha \longleftarrow  \alpha_{0}$ and $\beta \longleftarrow \beta_{0} $.
	\For {$i = 1$ \textbf{to} $t$}
		\State Applying \textbf{Algorithm \ref{Main Sequential decision Algorithm}}, decide if $MISO_{k}(T) \leq\ \frac{\alpha +\beta}{2}$.
		\If{ $MISO_{k}(T) \leq\ \frac{\alpha + \beta}{2}$}
			\State $\beta \longleftarrow \frac{\alpha + \beta}{2}$.
		\Else
			\State $\alpha \longleftarrow \frac{\alpha + \beta}{2}$.
		\EndIf
	\EndFor
	
	\State Let $\mathcal{A}$ be the $k$-subpartition output of Algorithm \ref{Main Sequential decision Algorithm} for deciding 		$MISO_{k}(T) \leq \beta$.
	\State \textbf{return} $MISO_{k}(T) = \cost(\mathcal{A})$ and $\mathcal{A}$.
\end{algorithmic}
\end{algorithm}

According to the procedures, the main steps of the sequential data clustering algorithm, can be summarized as follows.

\begin{enumerate}
\item Given the dataset of vectors $X$, construct the affinity graph $G$ on $X$ along with the edge weights $d(x_{i},x_{j}) = \|x_{i}-x_{j}\|_{2}$.\label{first Part}
\\
\item Find a minimum spanning tree $T$ for $(G,d)$ and calculate the functions $\varphi$, $\omega$ and $p$ for $T$.
\\
\item Order the vertices of $T$ in BFS order and determine the parent for each vertex.
\\
\item Calculate $\varphi^{*}$, $\varphi_{*}$, $p^{*}$, $p_{*}$, $\omega^{*}$ and $\omega_{*}$.
\\
\item Apply Algorithm~\ref{Main Sequential Algorithm}, iteratively call Algorithm~\ref{Main Sequential decision Algorithm} and find $MISO_{k}(T)$ with the minimizing subpartition $\mathcal{A} \in D_{k}(X)$.
\end{enumerate}

\section{Structure of parallel clustering algorithm}

In this section, firstly we give a brief description of  NVIDIA CUDA architecture and programming model. Then we explain our data structures and some preliminaries. Finally, we propose our parallel data clustering algorithm in detail.

\subsection{GPGPU and CUDA programming model}

In last two decades, improving the performance of the computer processors is generally achieved by increasing the clock rate, however this improvement is restricted by the physical limitations. Therefore, a large amount of research have been concentrated on parallelism by using multi-processors (instead of one processor).

Today, the graphics processors (GPU) have been evolved into the powerful and flexible processors that are designed to perform calculations on large amount of independent data concurrently. 
General-purpose computing on graphics processing units, GPGPU, refers to the use of GPU for the mathematical and scientific computing tasks.\\

CUDA is a software development kit (SDK), which has been developed by NVIDIA Corporation, used for writing programs on their GPUs and allows for the execution of C functions on graphics card.

\begin{figure}[h!]
\centering
\includegraphics[scale=0.21]{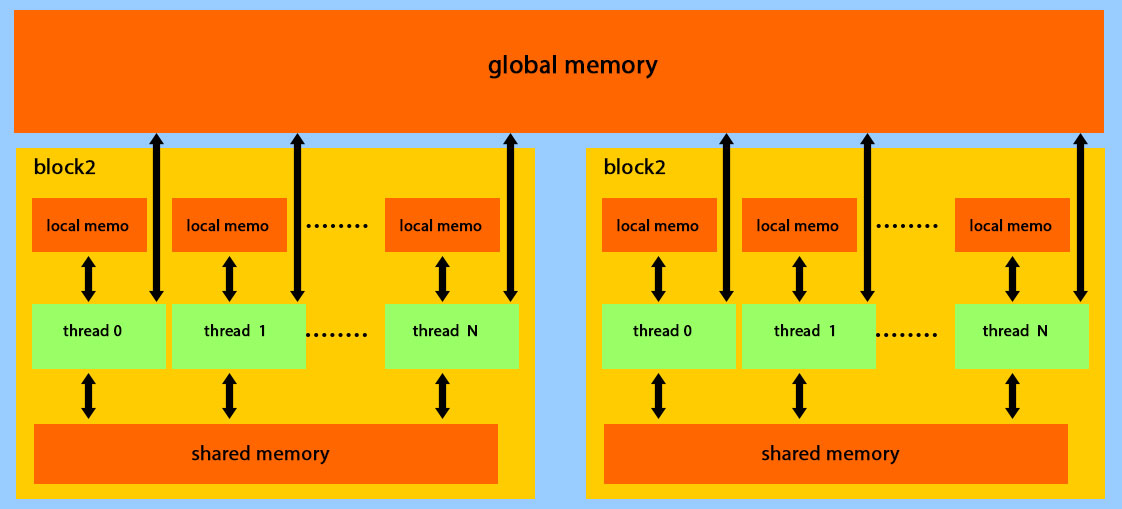}
    \caption{CUDA memory model}
    \label{cudamem}
    
\end{figure}

The CUDA programming model is based on executing parallel threads concurrently. CUDA manages threads in a hierarchical structure. A collection of threads (called a block) are run on a multiprocessor, which contains a number of processors, at a given time. The collection of all blocks in a single execution is called a grid. All threads in one grid share the same functionality, as they are being executed the same kernel code in multiple directions. Also, CUDA has a hierarchical memory model. In this model, each thread has a local memory. The threads of a single block share a fast-accessible memory which is called the shared memory. Also, all threads in different blocks can communicate through a low-speed global memory. This model provides a powerful platform for scientific computing objectives. (Figure \ref{cudamem})

\subsection{Principles and definitions}

The main essence of Algorithm \ref{Main Sequential decision Algorithm} is to check each vertex separately if it can be either cut from or join to its unique parent using the specific conditions. Thus, it is not hard to see that at each depth in the tree, all vertices can be considered independently and simultaneously. 
This is the crucial point in paralleling Algorithm \ref{Main Sequential decision Algorithm}. Based on this idea, we can describe our parallel data clustering algorithm as follows. 
\begin{enumerate}
	\item Given the dataset $X$, construct the affinity graph $(G,d)$ in parallel.
		\item Using Algorithm~\ref{Prim's Algorithm}, find the minimum spanning tree $T$ and the BFS order of the vertices.
		 \item Calculate the parameters $\varphi^{*}$, $\varphi_{*}$, $p^{*}$, $p_{*}$, $\omega^{*}$, and $\omega_{*}$.
			\item Start from the lowest depth and check the conditions of Algorithm~\ref{Main Sequential decision Algorithm}  for all vertices in that depth simultaneously.
\end{enumerate}


%
%

In the rest of this section, we describe each step of the algorithm in detail.

\subsection{Graph representations and data structures}
\begin{figure}
\centering
    \includegraphics[scale=0.25]{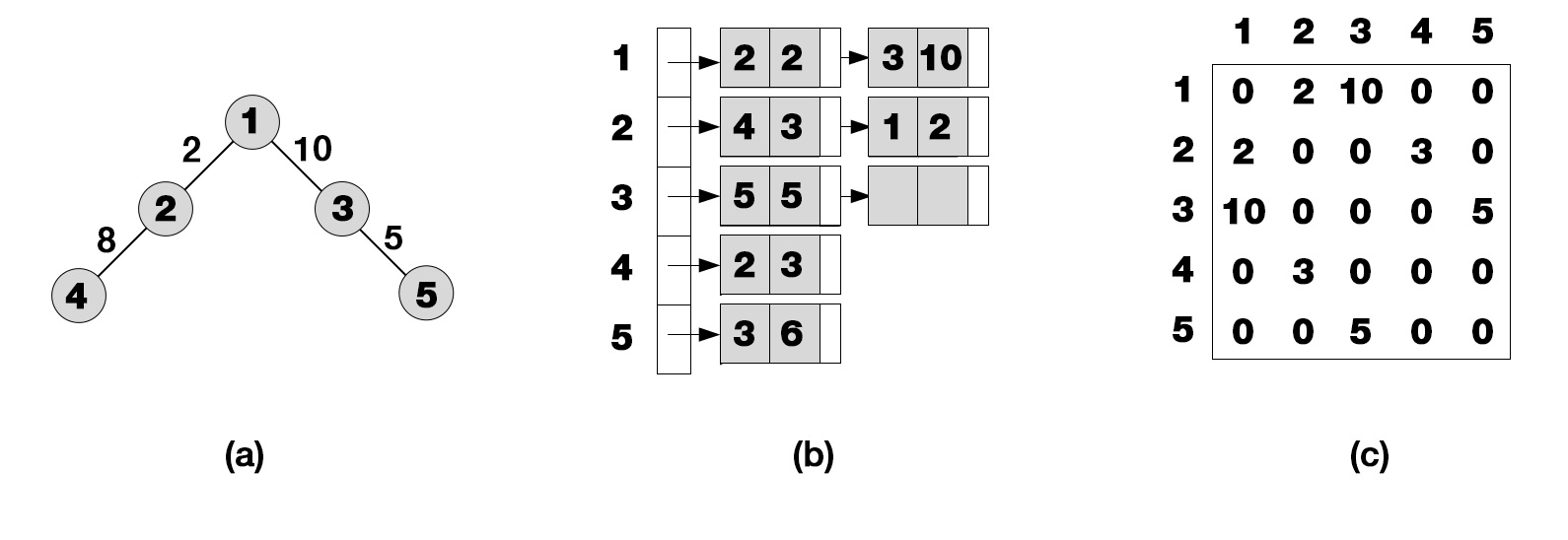}
    \caption{Standard graph representations.}
    \label{graphds}
    
\end{figure}

The standard data structure for graph representation are mainly the adjacency matrix and the adjacency list \cite{cormen2001introduction} (Figure~\ref{graphds}).\\

In the adjacency matrix of $G(V,E)$, we need $O(|V|^{2})$ space which is used for constructing an affinity matrix of the graph, where $G_{ij}$ shows the edge weight between two vertices $i$ and $j$. 
On the other hand, the adjacency list of a graph $G(V,E)$ consists of an array $A$ of lists of size $|V|$, one for each vertex in $V$ and for each $u \in V$ the adjacency list of $u$, $A[u]$ contain all neighbours  of $u$ and their edge weights.\\

\begin{figure}[h!]
    \centering
\includegraphics[scale=0.25]{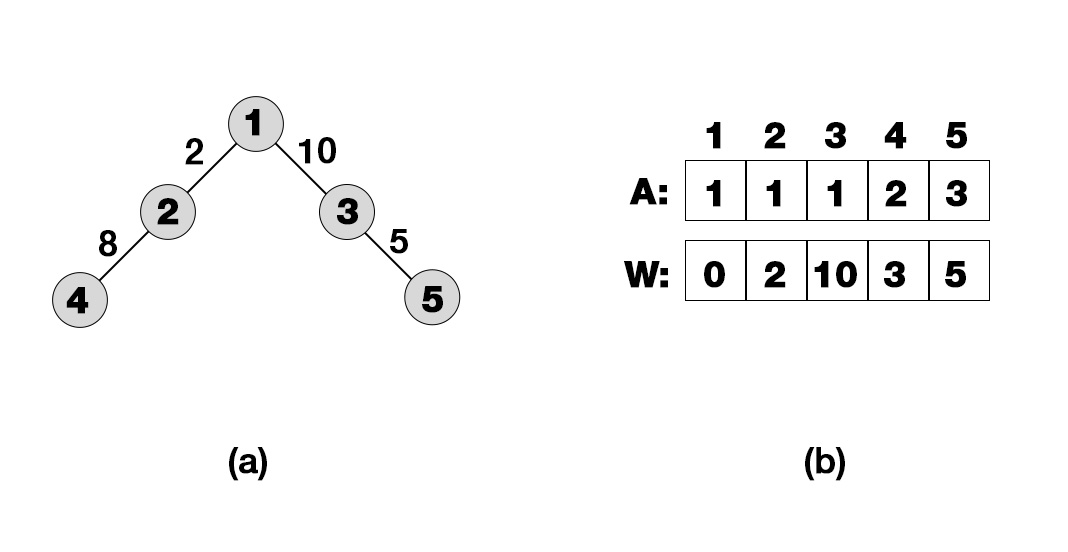}
    \caption{Parent List.}
    \label{weightlist}
    
\end{figure}

In our implementation, we use the adjacency matrix for constructing fully connected graph in order to access every edge weight for calculating the minimum spanning tree in $O(1)$. 
Also, we used a new data structure to represent the rooted tree called  \textit{parent list} which consists of two arrays $A$ and $W$ with size $|V|$, and for each vertex $u \in V(G)$, $A[u]$ contains the parent of $u$ and $W[u]$ saves the edge weight $\varphi({uv})$, where $v$ is the unique parent of $u$ (Figure \ref{weightlist}).

This data structure is more useful in the way that it is linear space bounded and the edge weights can be obtained in the constant time.
 
\subsection{Finding affinity matrix}
The parallel implementation for calculating the affinity matrix is one of the easiest steps in our algorithm. For calculating the distances of the pairs is totally independent, for each pair $(i,j)$ of vertices, one thread computes the corresponding weight $d(i,j)$ independently and the total number of launched threads is $ \frac{n(n+1)}{2}$. 


\subsection{Finding the minimum spanning tree and the weight functions}
Finding the minimum spanning tree (MST) is one of the most important and most studied problems in combinatorial optimization. It has many applications in different fields such as networking, VLSI layout and routing,  computational biology and etc.

Formally, a minimum spanning tree of an undirected connected graph $G(V,E)$ with vertices $V=\lbrace v_{1},\ldots,v_{n}\rbrace$ and weighed edges $E$, can be defined as a connected subgraph of $G$ that contains all vertices and $n-1$ edges with the minimum total weights.


The most well-known sequential MST algorithms are Kruskal, Prim and Boruvka algorithm \cite{cormen2001introduction}.
In \cite{wang2011design}, Wang et al. showed that compared to the other parallel MST algorithms, parallel implementation of Prim's algorithm is incomparably slower. This is due to the fact that in Prim's algorithm, the vertices are added to the tree sequentially. Nevertheless, Prim's algorithm is more useful for our purpose in the way that by adding vertices sequentially, we can determine some information including the parent and depth functions at the same time for each vertex. This information can lead us to omit the BFS  part of the algorithm.

In addition, Wang et al. \cite{wang2011design} proposed a parallel version of the Prim's algorithm and showed that their algorithm can be implemented efficiently on GPUs using min-reduction primitive (for more information about the parallel implementation of reduction algorithms see \cite{harris2007optimizing}).

Based on their approach, our parallel implementation of Prim's algorithm can described as follows:

%
%
%
%

\begin{algorithm}
\caption{\\
\textbf{Input:} A weighted complete graph $(G,d)$ and a root vertex $v$.\\
\textbf{Output:} The minimum spanning tree rooted at $v$ and the parent and depth functions.}\label{Prim's Algorithm}
\begin{algorithmic}[1]
\State Initialize  $T=\lbrace \emptyset \rbrace$.
\State Add the root vertex $v$ to the tree and save the distance between $v$ and the other vertices.
\While{$|T|\leq n$}
	\State Find the vertex outside $T$ with the minimum distance to $T$ using min-reduction algorithm.
	\State Add new vertex to $T$ and save its \textbf{Parent} and \textbf{Depth}.
	\State Update all distances simultaneously.
\EndWhile
\end{algorithmic}
\end{algorithm}

After calculating the minimum spanning tree, we apply a min-reduction algorithm to find the numbers $\varphi_{*}$,  $p_{*}$, $\omega_{*}$ and sum-reduction algorithm to calculate $\varphi^{*}$,  $p^{*}$, $\omega^{*}$.

\subsection{Computing  the isoperimetric number}

One may observe that Algorithm~\ref{Main Sequential decision Algorithm} is a search algorithm which starts by traversing vertices in the BFS order and decides to join or separate the vertices from their parents. Finally, it returns the suitable clusters (if exist). 
A vertex which is separated from its parent due to the conditions of Algorithm~\ref{Main Sequential decision Algorithm} is called a \textit{cut vertex}. We define two arrays $Cut$ and $\eta$ of size $n$ as follows. For every $i\in\{1,\ldots, n\}$, define
\[Cut(i)=\left\{ 
\begin{array}{ll}
1 & \text{ if } x_i \text{ is a cut vertex},\\
0& \text{otherwise}.
\end{array}\right.
\]
Also, at the end of Algorithm~\ref{Main Sequential decision Algorithm}, if $x_j$ is the vertex of the lowest depth which is joint with $x_i$, then define $\eta(i)=j$.
The arrays $Cut$ and $\eta$ can be computed after checking the condition during the execution of Algorithm~\ref{Main Sequential decision Algorithm}.

\begin{figure}[H]
\centering
\includegraphics[scale=0.445]{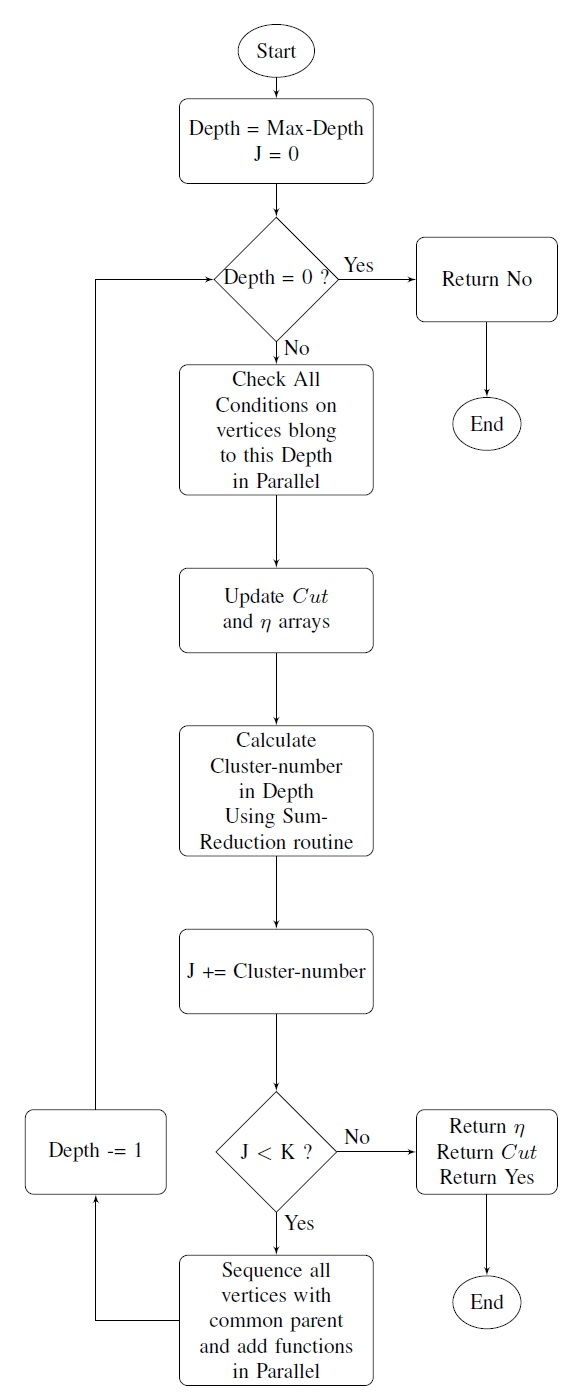}
\caption{Overall design flow diagram of the parallel implementation of  Algorithm \ref{Main Sequential decision Algorithm}.}\label{FlowChart}
\end{figure}

As mentioned above, in the parallel version, we check the conditions of Algorithm~\ref{Main Sequential decision Algorithm} depth-by-depth.
%
Figure~\ref{FlowChart} shows the overall design flow diagram.
Also, during the CUDA implementation, we use some techniques which significantly improve the performance of the program.
In the following, we will describe the details of these techniques.

\begin{enumerate}
	\item Since it is impossible to write on a single memory location in parallel at the same time, after checking the conditions of  Algorithm~\ref{Main Sequential decision Algorithm} at each depth, we try to sequence all vertices belonging to this depth with a common parent. We save this sequence for the whole vertices in an array called $ChildId$ which can be obtained during MST computation step (in Algorithm \ref{Prim's Algorithm}). Then, we use this array to prevent the conflict between the threads. Therefore, we can operate all vertices with the same $ChildId$ in parallel.
	\item Writing step is the most expensive part in the algorithm and based on this fact, we try to exchange the algorithm steps to reduce the execution number of this task. For this purpose, after checking the conditions, firstly, we calculate the number of clusters that can be obtained and  just execute the adding step if we have not achieved $k$ clusters.
	\item The final output of the data clustering algorithms is an array in size of the data containing a cluster label for each single element. Since, this array is just needed in the final step, we do not calculate these labels in the intermediate steps. Instead, at each step, we just save the arrays $\eta$ and $Cut$ (defined above) and we use them to produce the final labels, at the end of the Algorithm~\ref{Main Sequential Algorithm}. 

	\item Having two arrays $\eta$ and $Cut$ is sufficient to obtain the cluster labels for all elements. We designed an efficient procedure to calculate these labels by applying the parallel exclusive scan operation on $Cut$ array in the last step of Algorithm \ref{Main Sequential Algorithm}.
	(For more information about the parallel implementation of the exclusive scan algorithm see \cite{harris2007parallel}.)
\end{enumerate}

%
%
%
%
%

\newpage

Based on these facts, in Algorithm~\ref{Main Algorithm}, we provide a parallel implementation of Algorithm~\ref{Main Sequential Algorithm}.

\begin{algorithm}[H]
\caption{\\
\textbf{Input:} A weighted tree ($T,\omega,\varphi,p$) and an integer $k$. \\
\textbf{Output:} A minimizer achieving $MISO_{k}(T)$.}\label{Main Algorithm}
\begin{algorithmic}
	\State Let $\alpha_{0} \longleftarrow \frac{\varphi_{*} + p_{*}}{\omega^{*}}$ and $\beta_{0} \longleftarrow \frac{\varphi^{*} + p^{*}}{\omega_{*}}$.
	\State Let $t \longleftarrow \log(2\omega^{*^{2}}(\beta_{0} - \alpha_{0}))- \log(\varphi_{*} + p_{*} )$.
	\State Initialize $\alpha \longleftarrow  \alpha_{0}$ and $\beta \longleftarrow \beta_{0} $
	\For {$i = 1$ \textbf{to} $t$}
		\State Apply the parallel algorithm and decide if $MISO_{k} \leq\ \frac{\alpha + \beta}{2}$.
		\If{ $MISO_{k} \leq\ \frac{\alpha + \beta}{2}$}
			\State Save  $Cut$ and  $\eta$;
			\State $\beta \longleftarrow \frac{\alpha + \beta}{2}$
		\Else
			\State $\alpha \longleftarrow \frac{\alpha + \beta}{2}$
		\EndIf
	\EndFor

\State Initialize the array $\psi$ by $\psi(j) =0$ and the set $A_j$ by $A_j=\emptyset $,  for every $1\leq j \leq k$.
\State Apply \textbf{Exclusive Scan} algorithm on $Cut$ and save in $ScanArray$.
	
	\For{all $1\leq i\leq n$ in \textbf{parallel} }
		\If{$Cut({i}) = 1$}	
		\State $\psi(ScanArray({i})) = {i};$
		\EndIf
		\EndFor

	\For{$j = 1$ \textbf{to} $k$ }
		\For{all $1\leq {i}\leq n$ in \textbf{parallel} }
		\If{$\eta({i}) = \psi(j)$}
			\State $A_j  = A_{j} \cup \{ x_{i}\};$
		\EndIf		
	\EndFor
\EndFor

\State \Return $MISO_{k}(T)$ and $\mathcal{A}=\{A_1,\ldots, A_k\}$. 
\end{algorithmic}
\end{algorithm}


\section{Analysis and experimental results}

In this section, at first, we investigate the time complexity analysis of the algorithm and then we illustrate the performance of our algorithm on various synthetic and real datasets.

\subsection{Time Complexity Analysis}

According to the previous sections, the phases of our algorithm can be summarized as follows.

\begin{enumerate}

	\item \textit{ Adjacency matrix computation.} \\
		In the sequential algorithm, the computation of the adjacency matrix in global scale needs $O(n^{2})$ time to calculate all the edge weights. In our implementation each weight is calculated by a single thread. Thus, the time complexity is $O(1)$.
	\item \textit{Finding the minimum spanning Tree $T$.}\\
The running time of Prim's algorithm is $O(n^{2})$ in the  sequential implementation \cite{cormen2001introduction}. 
In our implementation, we begin with an empty tree and at each step, we use the min-reduction algorithm to find the nearest vertex. The step-complexity of the min-reduction algorithm is known to be $O(\log n)$ \cite{harris2007optimizing}. Therefore, the parallel implementation of Prim's Algorithm is $O(n \log n)$ \cite{wang2011design}.
	\item \textit{Tree Partitioning}\\
In \cite{daneshgar2013clustering}, it is shown that the time complexity of Algorithm \ref{Main Sequential Algorithm} is $O(n\log n)$.
In our parallel implementation of Algorithm~\ref{Main Sequential decision Algorithm}, we traverse the tree depth-by depth. At each depth firstly, we check all the conditions simultaneously which takes $O(1)$ time. Then, we sequence all vertices with a common parent and execute the operations.  Thus, the running time of the parallel algorithm is $O(n)$  and the time complexity of parallel tree partitioning algorithm is $O(n\log n)$ in the worst case.\\ It should be noted that the worst time occurs when all vertices are connected directly to the root and this case is too far from the real instances.
\end{enumerate}
\subsection{Experimental Results}
In this section, we compare the running time of our GPU and CPU implementations of the clustering algorithm. The computer used is equipped with Intel i7 3GHz CPU. It runs Linux Ubuntu and has 12GB of main memory. The GPU used is the NVIDIA GTX850M with 640 processing cores and 4 GB device memory and CUDA SDK 5.0 is used in our GPU implementation.

Table \ref{tab:Real Datasets} lists the datasets information and shows the performance of our Parallel implementation (GPU) with respect to the sequential implementation (CPU). 

\begin{table*}[htbp]
\centering
\caption{\bf Performance on UCI database. Symbols n,k,d  stand for the size of dataset, number of clusters and number of attributes (i.e. dimension).}
\scalebox{.8}{
\begin{tabular}{cccccccc}
\hline
Data Set & $ n $ & $k$ & $ d $ & CPU(ms) & GPU (ms) & Speed-Up & Misclassification rate \\
\hline 
Wine & 178 & 3 & 13 & 12.51 & 58.40 & 0.21x & 0.281\\
Skin  & 2000 & 2 & 3 & 373 & 311 & 1.19x &  0.014\\
Yeast & 1484 & 10 & 8 & 323 & 237 & 1.36x & 0.686\\
Wine Quality & 4898 & 10 & 11 & 4161 & 743 & 5.6x & 0.844 \\
Internet Advertisements & 3279 & 2 & 1558 & 177291 & 8483 & 20.89x & 0.482\\
\hline
\end{tabular}}
  \label{tab:Real Datasets}
\end{table*}

\begin{figure} [H]
\begin{center}

\includegraphics[scale=0.5]{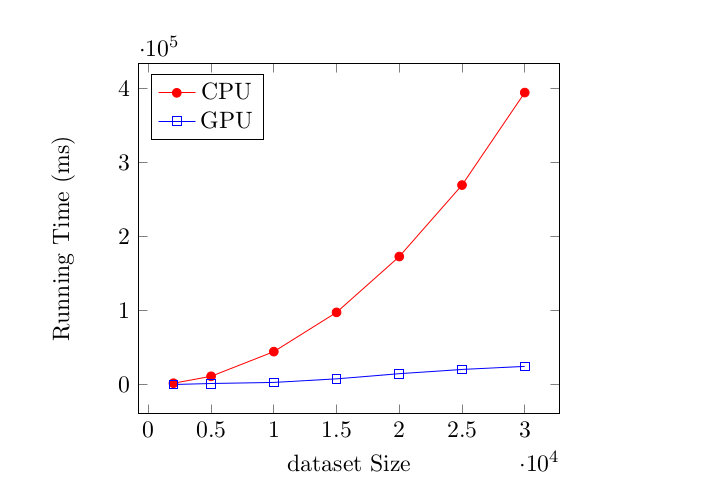}

\caption{Experiments on randomly generated data sets with high-dimensional data $(d=40)$.}\label{high dimension datasets}

\end{center}
\end{figure}

\begin{figure} [H]
\begin{center}
\includegraphics[scale=0.5]{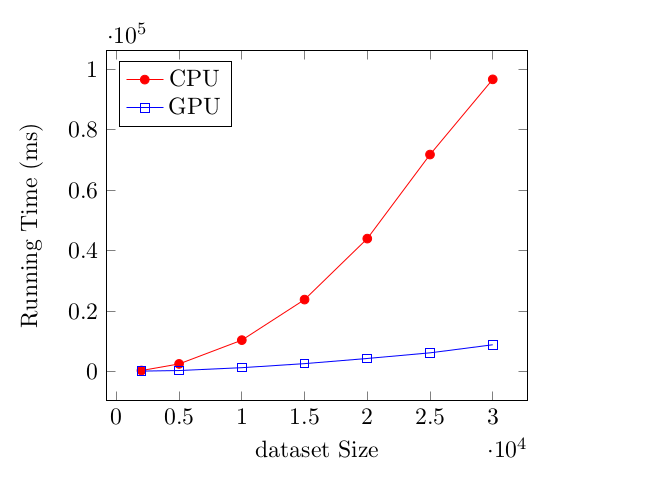}
\caption{Experiments on randomly generated datasets with low-dimensional data.}\label{low dimension datasets}

\end{center}
\end{figure}

Also, we have considered the performance of our algorithm on a number of randomly generated problems in low and high dimensions. Figures~\ref{high dimension datasets} and \ref{low dimension datasets} shows the performance for the data of high dimension ($d=40$) and low dimension ($d=5$), respectively.

\section{Conclusion And Future Work}
In this paper, we presented an efficient parallel data clustering algorithm based on the isoperimetric number of trees on GPU using CUDA. Firstly, we showed how to parallelize the procedure of computing adjacency matrix, the minimum spanning tree, and other preprocesses. Then, we described our parallel algorithm for tree partitioning and some tricks to increase its performance. Experiments show the efficiency of our algorithm  in the way of speed and accuracy. 

It should be noted that our algorithm is faced with the limitation of memory for large scale data sets on GPUs.
 So, it seems necessary to design a procedure to overcome this problem in the future work.

\bibliographystyle{elsarticle-num} 
\bibliography{ref}

\end{document}